\documentstyle[prl,aps,twocolumn,epsfig,floats]{revtex}
\begin{document}
\draft
\title{Observation of Quantum Asymmetry in an Aharonov-Bohm Ring}
\author{S. Pedersen, A.E. Hansen, A. Kristensen, C.B. S\o rensen, and P.E. Lindelof}
\address{The Niels Bohr Institute, University of Copenhagen, Universitetsparken 5, DK-2100 Copenhagen, Denmark}
\date{\today}
\maketitle
\begin{abstract}
We have investigated the Aharonov-Bohm effect in a one-dimensional GaAs/GaAlAs ring at low magnetic fields. The oscillatory magnetoconductance of these systems are for the first time systematically studied as a function of density. We observe phase-shifts of $\pi$ in the magnetoconductance oscillations, and halving of the fundamental $h/e$ period, as the density is varied. Theoretically we find agreement with the experiment, by introducing an asymmetry between the two arms of the ring. 
\end{abstract}

\section{Introduction}
The Aharonov-Bohm effect, first proposed in 1957 \cite{ab}, was experimentally realized in mesoscopic physics in 1987 \cite{timp1}. Soon after the Aharonov-Bohm effect became a very fruitful research area in mesoscopic physics \cite{timpagain,ford,ismail}. These early investigations, all except one \cite{chris}, focus their attention on the Aharonov-Bohm effect at relatively high magnetic fields ($\omega_{c} \tau\sim1$).

Recently, due to the perfection of device fabrication,  the Aharonov-Bohm effect has gained renewed interest.  Aharonov-Bohm rings are now used to perform phase sensitive measurement  on e.g.\ quantum dots \cite{heiblum} or on rings were a local gate only affects the properties in one of arms of the ring \cite{mailly}. The technique in these reports use the idea, that by locally changing the properties of one of the arms in the ring, and studying the Aharonov-Bohm effect as a function of this perturbation, information about the changes in the phase can be subtracted from the measurements. Recently also a realisation of the electronic double slit interference experiment presented surprising results \cite{yacoby}. Especially the observation of a period halving from $h/e$ to $h/2e$ and phase-shifts of $\pi$ has attracted large interest in these reports.

All these recent investigations are, as in contrast to the prior ones, performed at relative low magnetic fields and the perturbation enforced on the ring is regarded as local. Furthermore they are all performed in the multi-mode regime. Hence we find it of importance to study  the Aharonov-Bohm effect in the single-mode regime at low magnetic fields and as a function of a global perturbation.

\section{experiment}
Our starting point in the fabrication of the Aharonov-Bohm structures is a standard two dimensional electron gas (2DEG) realized  in a GaAs/GaAlAs heterostructure.  The two dimensional electron density is $\rm{n}=2.0 \cdot 10^{15}\rm{m}^{-2}$ and the mobility of the heterostructure is $\mu = 90\rm{T}^{-1}$. This corresponds to a mean free path of approximately $6 \rm{\mu m}$. The 2DEG is made by conventional molecular beam epitaxy (MBE) and is situated 90nm below the surface of the heterostructure. For further details regarding the wafer, contacts etc. we refer to \cite{Anders}. 

\begin{figure}
\begin{center}
\epsfig{file=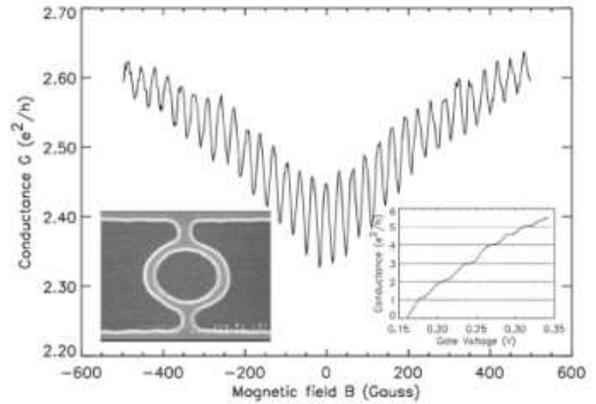,width=0.5\textwidth}
\end{center}
\caption{\small{Measured magnetoconductance of the device shown on the SEM picture in the left insert. The magnetoconductance show a very clear Aharonov-Bohm signal imposed on a slowly varying background. The right insert displays the zero magnetic field conductance at $4.2 {\rm K}$ as a function of gate voltage. The conductance curve displays distinct steps which show that the device is in a single- or few-mode regime, see text.}\protect}
 \label{1}
 \end{figure}

Using a rebuild field emission scanning electron microscope (SEM) operated at an acceleration voltage of $30\rm{keV}$ a $100\rm{nm}$ thick PMMA etch mask is defined by standard e-beam lithography (EBL) on the surface of the heterostructure. The pattern written in the PMMA was then transferred to the by a $\rm{50nm}$ shallow etching in $\rm{H_{3}PO_{4}:H_{2}O_{2}:H_{2}O}$. The dimensions of the etched Aharonov-Bohm structure is given by a ring radius $r=0.65 \rm{\mu m}$ and a width of the arms $w=200\rm{nm}$ as can be seen on Fig.\ref{1}. 

In a second EBL step we define a PMMA lift-off mask for a $\rm{50 nm}$ thick and $30\rm{\mu m}$ wide gold gate which covers the entire Aharonov-Bohm ring. This allows us to globally control the electron density in the Aharonov-Bohm ring during the measurements. Due to depletion from the edges, the structure is initially pinched off. By applying a positive voltage $V_{g}$ on the global gate, electrons are accumulated in the structure and the structure begins to conduct.

The sample was mounted on a $^{3}$He cryostat, equipped with a copper electromagnet. All measurements are performed at $0.3 {\rm K}$ if nothing else is mentioned.
The measurements was performed by a conventional voltage biased lock-in technique with an excitation voltage of $V_{\rm{pp}}=7.7 {\rm \mu V}$ at a frequency of $131 {\rm Hz}$. In this report we focus on measurements performed on one device, almost identical results have been obtain with another device in a total of six different cool-downs.

\section{results and discussion}
Fig.\ref{1}  present a measurement of the magnetoconductance of the device displayed in the left insert. As expected the magnetoconductance is dominated by the Aharonov-Bohm oscillations. The measurement is, due to the long distance between the voltage probes, an effectively two-terminal measurement; hence the Aharonov-Bohm magnetoconductance is as observed forced to be symmetrical due to the Onsager relations. 

A Fourier transform of the magnetoresistance displays a very large peak corresponding to a period of 33Gs.  This is in full agreement with the dimensions of the ring obtained from the SEM picture.

The right insert i Fig.\ref{1} displays the conductance as a function of gate voltage at $4.2 {\rm K}$, steps are observed at integer values of $e^{2}/h$. Five steps are seen as the voltage is increased with $0.18{\rm V}$ from pinch-off. Such steps have previously been reported in Aharonov-Bohm rings \cite{ismail}, and can be interpreted as if the device, at these relatively high temperatures, behaves as two quantum point contacts (QPC) in series  with two QPC in parallel. In any case, this indicates that our system, in the gate voltage regime used here, only has a few propagating modes.
When the temperature is lowered, the conductance curve changes into a fluctuating signal, and finally the steps are completely washed out by the fluctuations. 
These fluctuations, which we ascribe to resonance's, appear simultaneously with the Aharonov-Bohm oscillations and are the signature of a fully phase coherent device.

\begin{figure}
\begin{center}
\centerline{
\epsfig{file=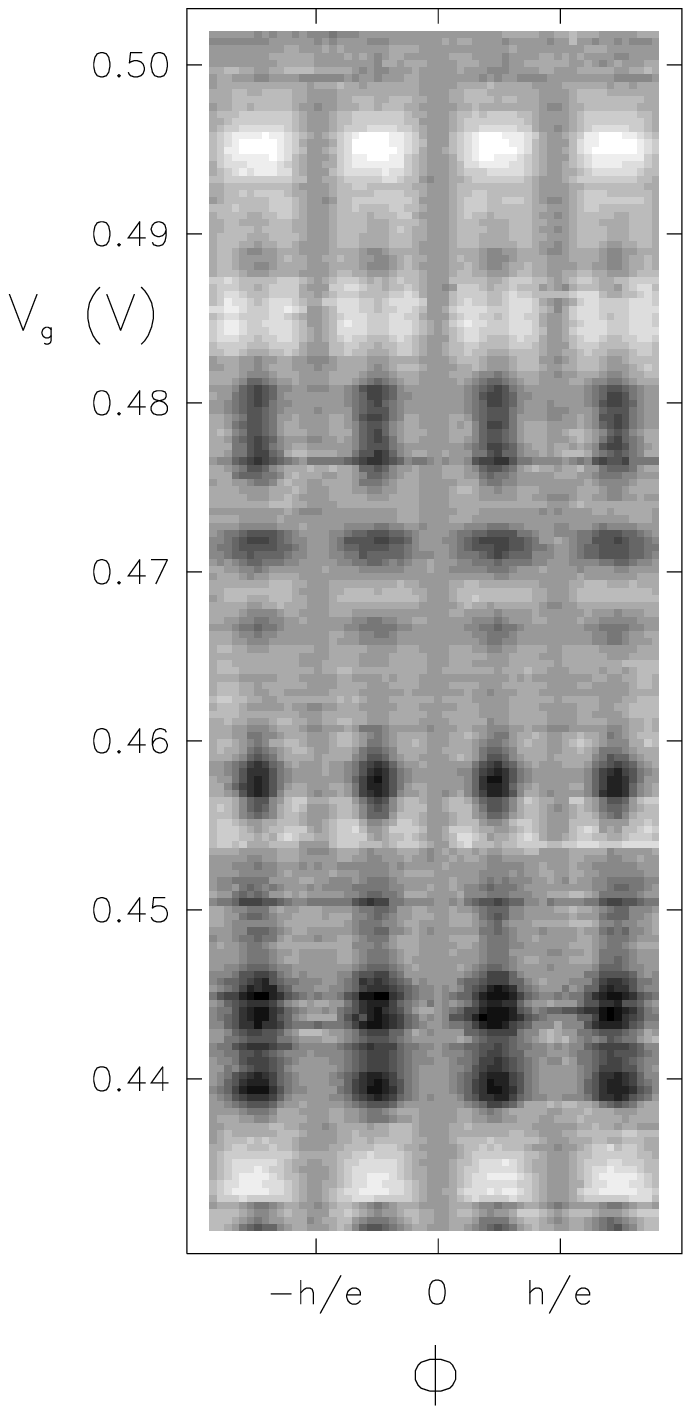,width=0.25\textwidth}
\epsfig{file=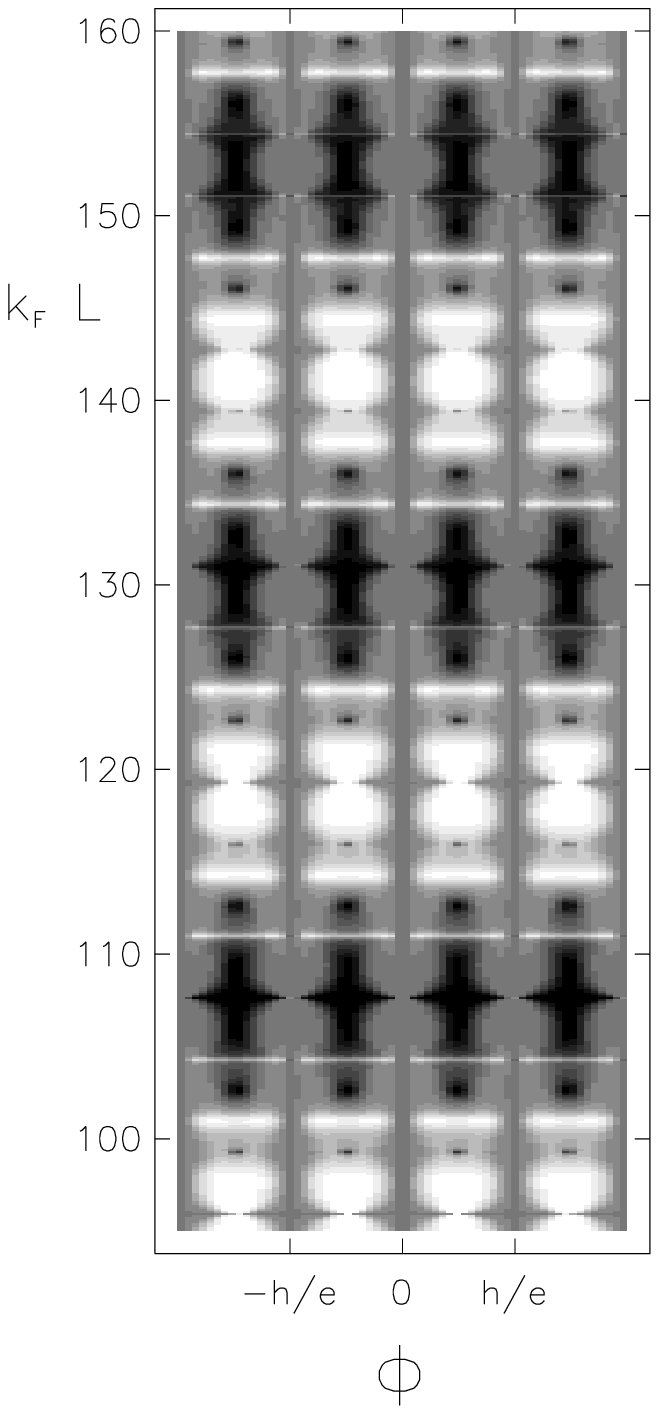,width=0.25\textwidth}
}
\end{center}
 \caption{\small{Left insert: The measured Aharonov-Bohm signal viz. $G(B,V_{G})-G(0,V_{G})$, as a function of applied magnetic flux $\phi$ through the ring(horizontal axis) and global gate voltage $V_{g}$(vertical axis). The right insert displays the calculated magnetoconductance as a function of magnetic flux $\phi$ (horzontal axis) and ${\rm k_{f}L}$ (vertical axis). The dark (white) regions corresponds to minima (maxima).}\protect}
 \label{2}
 \end{figure}

Fig.\ref{2} shows two contour plots; the left one displays the measured conductance $G(B,V_{g})$ as a function of gate voltage and magnetic field. The fluctuating zero magnetic field conductance $G(0,V_{g})$ has been subtracted from the measurements to enhance the contours of the Aharonov-Bohm signal. 
This figure clearly displays the Aharonov-Bohm oscillations as a periodic pattern in the horizontal direction. In the vertical direction the gate voltage is changed between $0.43 {\rm V}$ and $0.50 {\rm V}$.  
The studied systems begins to conduct at $0.33 {\rm V}$. The data clearly shows that, by changing the gate voltage viz. changing the density, it is possible to change the sign of the magnetoconductance - or stated differently, change the phase of the Aharonov-Bohm signal by $\pi$ . This is quite surprising since a negative magnetoconductance always are expected in symmetrical Aharonov-Bohm structures \cite{bil1}. However in the case of an asymmetrical structure this is no longer the case \cite{bil2}.

In order to compare our measurements with theory \cite{bil2}, we need to estimate the electron density ${\rm n}$ in the device. From a simple capacitor based estimate a voltage of $0.5 {\rm V}$ corresponds to a electron density of ${\rm n}=\epsilon (0.50{\rm V}-0.33{\rm V})/ae=1.36 \cdot 10^{15} {\rm m}^{-2}$. Here $a=90{\rm nm}$ is the distance from the wafer surface to the 2DEG.
At high magnetic fields we observe the so called Camel-back structure \cite{beenakker}. An analysis of this structure yields a density of ${\rm n}=0.95 \cdot 10^{15} {\rm m}^{-2}$ at the gate voltage $V_{g}=0.50V$. We therefore estimate the electron density at $0.50 {\rm V}$ to be $1 \cdot 10^{15} {\rm m}^{-2}$ and to be zero at $0.33 {\rm V}$. The characteristic dimensionless number ${\rm k_{f}L}$ is found to be approximately $160$ at $V_{g}=0.50V$, where $\rm{L}$ is half the circumference of the ring and $\rm{k_{f}}=\sqrt{2 \pi {\rm n}}$ is the Fermi wavevector.

As a first order approximation we can use a linear relation between  the Fermi wavevector and the gate voltage, viz.
\begin{equation}
\label{relation}
{\rm k_{f}L}=160 \frac{{\rm V_{g}}-0.33{\rm V}}{0.50{\rm V}-0.33{\rm V }}
\end{equation}

In the case of asymmetrical structures the conductance is given by \cite{bil2}
\begin{eqnarray}
\label{bil}
{\rm G}(\theta, \phi, \delta )=& & \frac{2e^{2}}{h} 2\epsilon {\rm g}(\theta,\phi) \nonumber\\
& &(\sin^{2}\phi\cos^{2}\theta+\sin^{2}\theta\sin^{2}\delta-\sin^{2}\phi\sin^{2}\delta),
\end{eqnarray}
where $\theta=\pi \Phi/\Phi_{0}$ is the phase originating from the magnetic flux,  $\phi=k_{f}L$ is the average phase due to spatial propagation, $\delta=\Delta(k_{f}L)$ is the phase difference between the two ways of traversing the ring. The coupling parameter $\epsilon$ can vary between $1/2$ for a fully transparent system and zero for a totally reflecting system. The function  ${\rm g(\theta, \phi)}$ is given by
\begin{eqnarray}
& &{\rm g(\theta, \phi)}= \nonumber\\
& &\frac{2 \epsilon}{(a_{-}^{2}\cos2\delta+a_{+}^{2}\cos2\theta-(1-\epsilon)\cos2\phi)^{2}+\epsilon^{2}\sin^{2}2\phi},
\end{eqnarray}
where $a_{\pm}=(1/2)(\sqrt{1-2\epsilon}\pm1)$.

The right part of Fig.\ref{2} shows a plot of equation (\ref{bil}), where the value of the phase due to asymmetry is set to vary as $\delta=0.15 \cdot {\rm k_{f}L}$ and the coupling parameter $\epsilon$ is set to $1/2$. The scale on the ${\rm k_{f}L}$-axis is determined by using equation (\ref{relation}), hence when the voltage is changed between $0.50{\rm V}$ and $0.43{\rm V}$ the value of ${\rm k_{f}L}$ changes between 160 and 95.
It is seen in Fig.\ref{2} that even though a perfect fit is not possible, it is indeed possible to reproduce the general features of the measurement by the theoretical expression (\ref{bil}).
\begin{figure}
\begin{center}
\epsfig{file=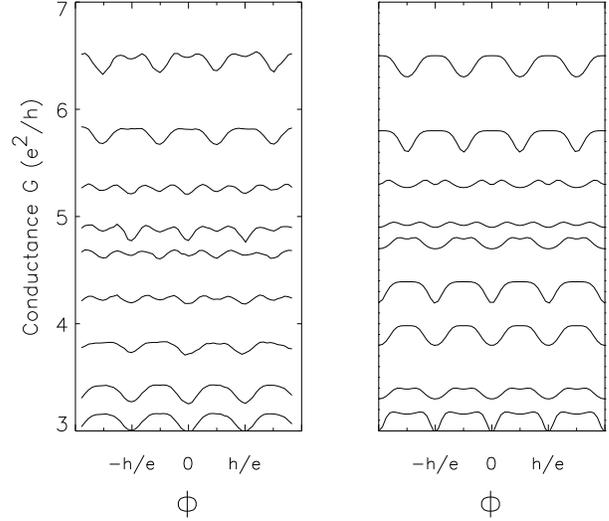,width=0.45\textwidth}
\end{center}
\caption{\small{Left insert: Equidistantly spaced ($\triangle V_{g}=2.2{\rm mV}$) measured magnetoconductance traces as a function of applied magnetic field ${\rm B}$. Right insert: Equidistantly spaced ($\triangle({\rm k_{f}L})=2.0$) magnetoconductance traces as a function of magnetic field ${\rm B}$ calculated from expression (\ref{bil}). Note that the major characteristics such as phase-shifts and period halving are observed in both inserts.}\protect}
 \label{3}
 \end{figure}
Fig.\ref{3} shows traces taken from the contour plots of Fig.\ref{2}. The figure on the left shows nine equidistantly spaced measurements $(\triangle V_{g}=2.2 {\rm mV})$. According to equation (\ref{relation}) this corresponds to a equidistant spacing of $2.0$ in units of ${\rm k_{f}L}$. On the figure to the right nine successive theoretical magnetoconductance curves are presented, the distance in ${\rm k_{f}L}$ is also $2.0$. The amplitude of all nine curves has been scaled by a factor of $0.2$. We ascribe this decrease in amplitude to the fact that the experiment is performed at finite temperatures whereas the theoretical expression is an effective zero temperature result.

From the comparison of the theoretical expression and the measured magnetoconductance curves it is seen that a direct comparison between single traces is possible in a limited voltage regime. Such a comparison is not possible over the whole gate voltage regime from $V_{g}=0.43{\rm V}$ to $V_{g}=0.50{\rm V}$ with the assumption of a linear relation between gate voltage and $k_{f}$ made here, see equation (\ref{relation}). One should also note, that when changing $V_{g}$ with $0.07 \rm{V}$ a new sublevel will start to get populated, as can be seen on the insert of Fig.\ref{1}. However, all the complicated features such as period halving and changes of $\pi$ in the phase of the Aharonov-Bohm signal are obseved for both theory and experiment.

\section{conclusion}
We have measured the Aharonov-Bohm effect in a one-dimensional GaAlAs/GaAs ring. The effect was studied as a function of the electron density in the ring. We find that the standard theoretical expressions for a symmetrical ring are not applicable for the rings in question. To reproduce essential features of the measurements, i.e. the phase shifts, it is necessary to introduce a in build asymmetry in the ring - e.g. different average density in the two arms of the ring.

These results for the first time show the influence of asymmetry of a ring and it gives insight in the recent observation of phase-shifts and period halving  seen in other related systems.

\section{acknowledgements}
This work was financially supported by Velux Fonden, Ib Henriksen Foundation, Novo Nordisk Foundation, Danish Research Council (grant 9502937, 9601677 and 9800243) and the Danish Technical Research Council (grant 9701490)


%
%


\begin{references}
\bibitem{ab} Y. Aharonov and D. Bohm, Phys. Rev, {\bf 115} (1959) 485.
\bibitem{timp1} G. Timp, A.M. Chang, J.E. Cunningham, T.Y. Chang, P. Mankiewich, R. Behringer, and R.E. Howard, Phys. Rev. Lett. {\bf 58} (1987) 2814.
\bibitem{timpagain} G. Timp et al., Phys. Rev. B {\bf39} (1989) 6227.
\bibitem{ford}C.J.B. Ford, T.J. Thornton, R. Newbury, M. Pepper, H. Ahmed, C.T. Foxon, J.J. Harris, J.Phys: Solid State Phys. {\bf 21} (1988) L325; C.J.B Ford, T.J. Thornton, R. Newbury, M. Pepper, H. Ahmed, D.C. Peacock, D.A. Ritchie, J.E.F. Frost, and G.A.C Jones, Appl. Phys. Lett {\bf 54} (1989) 21.
\bibitem{ismail} K. Ismail, S. Washburn, and K.Y. Lee, Appl. Phys Lett. {\bf 59} (1991) 1998; J. Liu, K. Ismail, K.Y. Lee, J.M. Hong, and S. Washburn, Phys. Rev B {\bf 47} (1993) 13039; ibid. {\bf 48} (1993) 15148; ibid. {\bf 50} (1994) 17383.
\bibitem{chris} C.J.B. Ford, A.B. Fowler, J.M. Hong, C.M. Knoedler, S.E. Laux, J.J. Wainer, and S. Washburn, Surf. Sci {\bf 229} (1990) 307.
\bibitem{heiblum}A. Yacoby, M. Heiblum, D. Mahalu, and Hadas Shtrikman. Phys. Rev. Lett. {\bf 74} (1995) 4047; A. Yacoby, R. Schuster, and M. Heiblum. Phys. Rev. B. {\bf 53} (1996) 9583; R. Schuster, E. Buks, M. Heiblum, D. Mahalu, V. Umansky, and Hadas Shtrikman. Nature {\bf 385} (1997) 417.
\bibitem{mailly} G. Cernicchiaro, T. Martin, K. Hasselbach, D. Mailly, and A. Benoit, Rev. Lett {\bf 79} (1997) 273.
\bibitem{yacoby}A. Yacoby, M. Heiblum, V. Umansky, H. Shtrikman, and D. Mahalu, Phys. Rev. Lett. {\bf 73} (1994) 3149.
\bibitem{Anders} A. Kristensen, P.E. Lindelof, J.B. Jensen, M.Zaffalon, J. Hollingbery, S. Pedersen, J. Nyg\aa rd, H. Bruus, S.M. Reimann, C.B. S\o rensen, M. Michel, and A. Forchel, Physica B {\bf 249-251} (1998)180-184; A. Kristensen, J.B. Jensen, M. Zaffalon, C.B. S\o rensen, S.M. Reimann, P.E. Lindelof, M. Michel, and A. Forchel, J. Appl. Phys {\bf 83} (1998) 607.
\bibitem{beenakker} C.W.J. Beenakker and H. van Houten, Phys. Rev. B. {\bf 39} (1989) 10445.
\bibitem{bil1} M.B\"uttiker, Y. Imry and M.Ya. Azbel, Phys. Rev. B {\bf 30} (1984) 1982; Y. Gefen, Y. Imry and M. Ya. Azbel, Phys Rev. Lett. {\bf 52} (1984) 129.
\bibitem{bil2}M. B\"uttiker, SQUID'85 - Superconducting Quantum Interference Devices and their Applications, edited by H.D. Haklbohm and H. L\"ubbig (Walter de Gruyter, Berlin, New York 1985) page 529. 
\end{references}
\end{document}